\documentclass[preprint, a4paper]{revtex4}
\usepackage[T1]{fontenc}
\usepackage{graphicx}
\usepackage{rotating}
\usepackage{bm}        
\usepackage{amssymb}   
\usepackage{mathrsfs}
\usepackage{amsmath}
\usepackage{bm}
\def\k1{k_1}
\def\k2{k_2}
\def\q1{q_1}
\def\q2{q_2}

\def\({\left (}
\def\){\right )}
\def\[{\left [}
\def\]{\right ]}

\newcommand{\beq}{\begin{equation}}
\newcommand{\eeq}{\end{equation}}

\begin{document}
\date{\today}
\flushbottom \draft
\title{
Cold collisions of complex polyatomic molecules}
\author{Zhiying Li and Eric J. Heller}
\affiliation{
Department of Physics, Harvard University, Cambridge, MA 02138, US
}
\begin{abstract}
We introduce a method for classical trajectory calculations to simulate collisions between atoms and large rigid asymmetric-top molecules. Using this method, we investigate the formation of molecule-helium complexes in buffer-gas cooling experiments at a temperature of 6.5~K for molecules as large as naphthalene. Our calculations show that the mean lifetime of the naphthalene-helium quasi-bound collision complex is not long enough for the formation of stable clusters under the experimental conditions. Our results suggest that it may be possible to improve the efficiency of the production of cold molecules in buffer-gas cooling experiments by increasing the density of helium. In addition, we find that the shape of molecules is important for the collision dynamics when the vibrational motion of molecules is frozen. For some molecules, it is even more crucial than the number of accessible degrees of freedom. This indicates that by selecting molecules with suitable shape for buffer-gas cooling, it may be possible to cool molecules with a very large number of degrees of freedom. 

\end{abstract}
\maketitle

\clearpage
\newpage
\section{Introduction}

Recent experimental work with ultracold molecules has generated a revolution in molecular physics \cite{Levi, Bethlem1, Julienne,  Krems1, Doyle,Krems2,Krems3}.  However, most of the experimental and theoretical studies in this field to date have focussed on diatomic or few-atom molecules \cite{Krems3, Heiner,Patterson1}.  An emerging direction in this field aims to extend the cooling techniques and study of ultracold diatomic molecules to large polyatomic molecules \cite{Kupper,Patterson2}. Cooling complex polyatomic molecules to cold and ultracold temperatures, if possible, will open up exciting opportunities for studying novel phenomena in physics and chemistry. For example, K\"upper {\it et al} have recently demonstrated how to manipulate the motion of polyatomic molecules in molecular beams using external fields \cite{Kupper}. This may lead to the development of new experiments on precision measurements of large molecule collisions. One can also use intense laser fields \cite{Stapelfeldt} or static electric fields \cite{Friedrich} to align cold molecules along a certain axis in the laboratory frame, which may provide a diversity of applications in molecular optics \cite{Lappas, Velotta, Kalosha, Bartels}, controlled chemistry \cite{Larsen}, and spectroscopic analysis \cite{Seideman1,Seideman2}. Cooling ensembles of polyatomic molecules is also necessary for the research of cold chemistry \cite{Krems3}.

Among all cooling techniques, buffer-gas cooling is the most powerful and versatile method that produces molecules in translationally and internally cold states \cite{Weinstein, Hancox, Maxwell}. This technique relies on energy thermalization of molecules in a buffer gas (e.g., helium) maintained at a cold temperature. Until recently, it was expected that polyatomic molecules, when placed in a cold buffer gas, will collect buffer gas atoms into clusters, which should make cooling polyatomic molecules difficult. However, Patterson {\it et al} have recently measured the thermalization rates of naphthalene (N) molecules in a cold helium gas \cite{Patterson2}. Contrary to the expectation, they observed no clustering. They proposed an explanation based on the Lindemann mechanism \cite{Patterson2}. Their model suggests that if the energy of the first excited vibrational state of the molecule is higher than the molecule-helium binding energy, there should be no clustering observed. However, this model excludes the role of rotational degrees of freedom in cold collisions. Could the rotation induce a collision complex and thus trapping-clustering via a Lindemann mechanism? 

Here we undertake a detailed study of the collision dynamics between helium and two kinds of molecules, naphthalene and benzene, under conditions close to buffer-gas cooling experiments. At temperatures below 10~K, these molecules reside in the ground vibrational state. We therefore treat the molecules as rigid. The rigid approximation can be justified semiclassically by noting that the vibrational modes in question are really ground state wavepackets represented by a mean position at rest. The conventional classical trajectory calculations for high energy dynamics allow all atoms in a molecule to move. \cite{Sewell}. These calculations, however, prove to be difficult to preserve the rigidity of molecules and are hard to implement for low temperature collisions excluding vibrational transitions. In addition, these methods meet with the zero point energy problem that may lead to unphysical energy localization in particular vibrational modes. Here, we adapt a molecular dynamics theory used in the simulation of polyatomic fluids to study low temperature collisions \cite{Evans1,Evans2,Rapaport}. In our approach, the number of differential equations to solve is independent of the number of atoms in the molecule. Only the center-of-mass of molecules participates in the equations of the relative motion in a space-fixed (SF) frame. The rotational motion of the molecules is described  separately by quaternion parameters in a body-fixed (BF) frame. We maintain the shape of the  molecules by converting the coordinates of atoms in the molecules from the BF frame to the SF frame by a rotation matrix constructed using the quaternion parameters. The rigidity of the molecules is therefore preserved without introducing internal coordinates to constrain the bond length and angles of the molecules. 

Low energy collisions have been an interesting research topic for a long time \cite{LaBudde, Augustin}. However, classical trajectory studies of collisions involving rigid molecules are restricted to small and symmetric-top molecules. This paper provides the first classical study of collision dynamics between a large rigid asymmetric-top molecule and an atom.  The method introduced here can also find many other applications.

\section{Methodology}

\begin{figure}[ht]
\begin{center}
\includegraphics[scale=0.6]{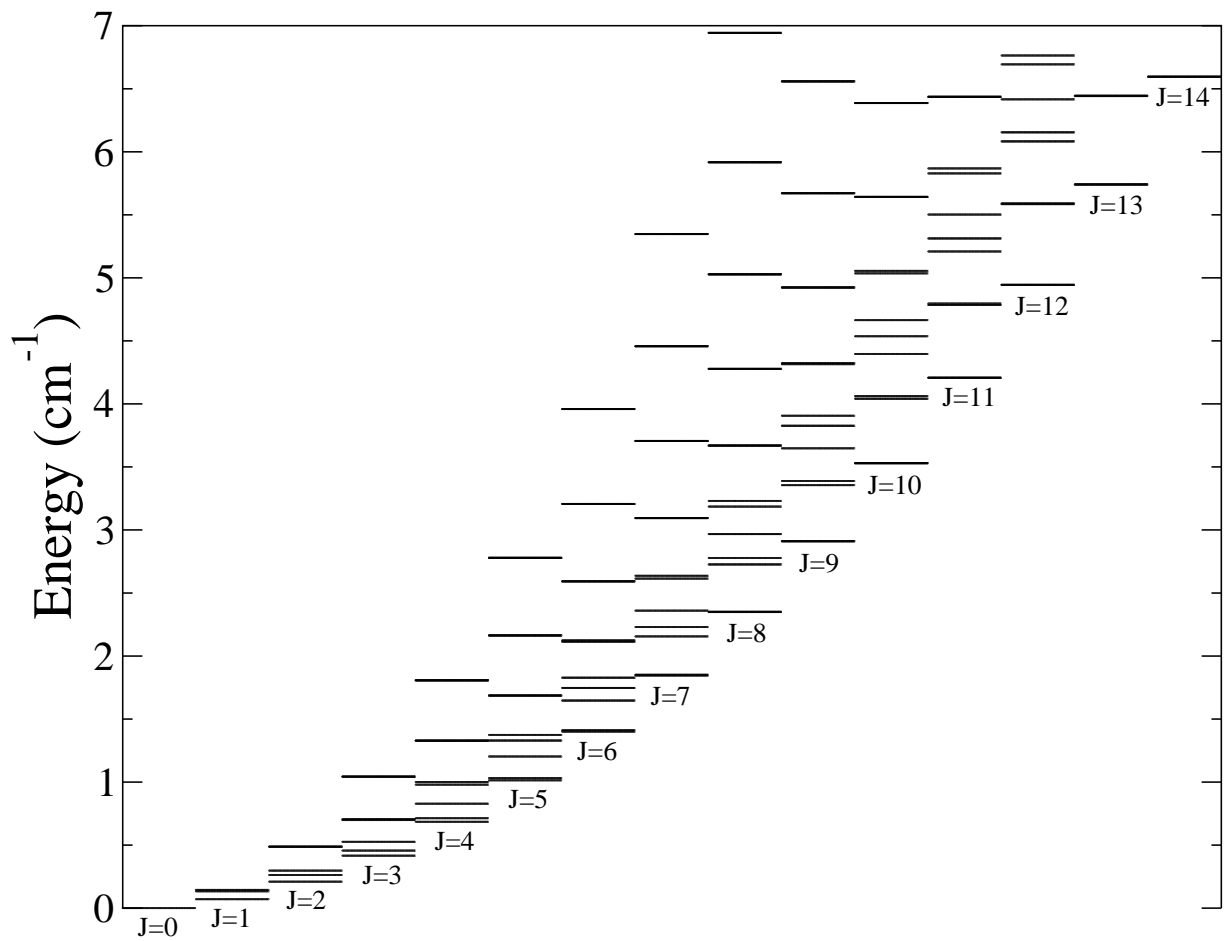}
\end{center}
\caption{The rotational energy levels of naphthalene below 7 cm$^{-1}$. }
\label{E-levels}
\end{figure}

We carry out classical trajectory calculations for naphthalene-helium and benzene-helium collisions at a temperature $T = 6.5$~K. In this section, we illustrate the methodology by naphthalene molecules. Naphthalene is a complex asymmetric-top molecule which has a large number of degrees of freedom. As shown in Figure.~\ref{E-levels}, there are over one hundred non-degenerate rotational energy levels below 7~cm$^{-1}$. The complexity of the naphthalene molecules precludes the fully quantum calculation of the He-naphthalene collision dynamics. At low temperatures relevant for the buffer-gas cooling experiments, however, particles may behave classically. The thermal de Broglie wavelength of helium at the temperature of 6.5~K is about 3.56~\AA. The mean interatomic distance in the buffer gas is around 136~\AA. This indicates that quantum effects in the collision dynamics are insignificant \cite{Annett}. Moreover, a typical classical action associated to energy transfer between helium and naphthalene at this temperature is on the order of  10~cm$^{-1}$. It is much larger than the rotational energy splitting of naphthalene, as shown in  Fig.~\ref{E-levels}, suggesting that the dynamics under experimental conditions falls into the classical regime. We therefore employed a classical dynamics approach. In order to calibrate the accuracy of our calculations, we calculate the total thermal cross section for N$\--$He collisions at $T = 6.5$~K. Our results are consistent with the experimental data. 

 \subsection{The system}

\begin{figure}[ht]
\begin{center}
\includegraphics[scale=0.5]{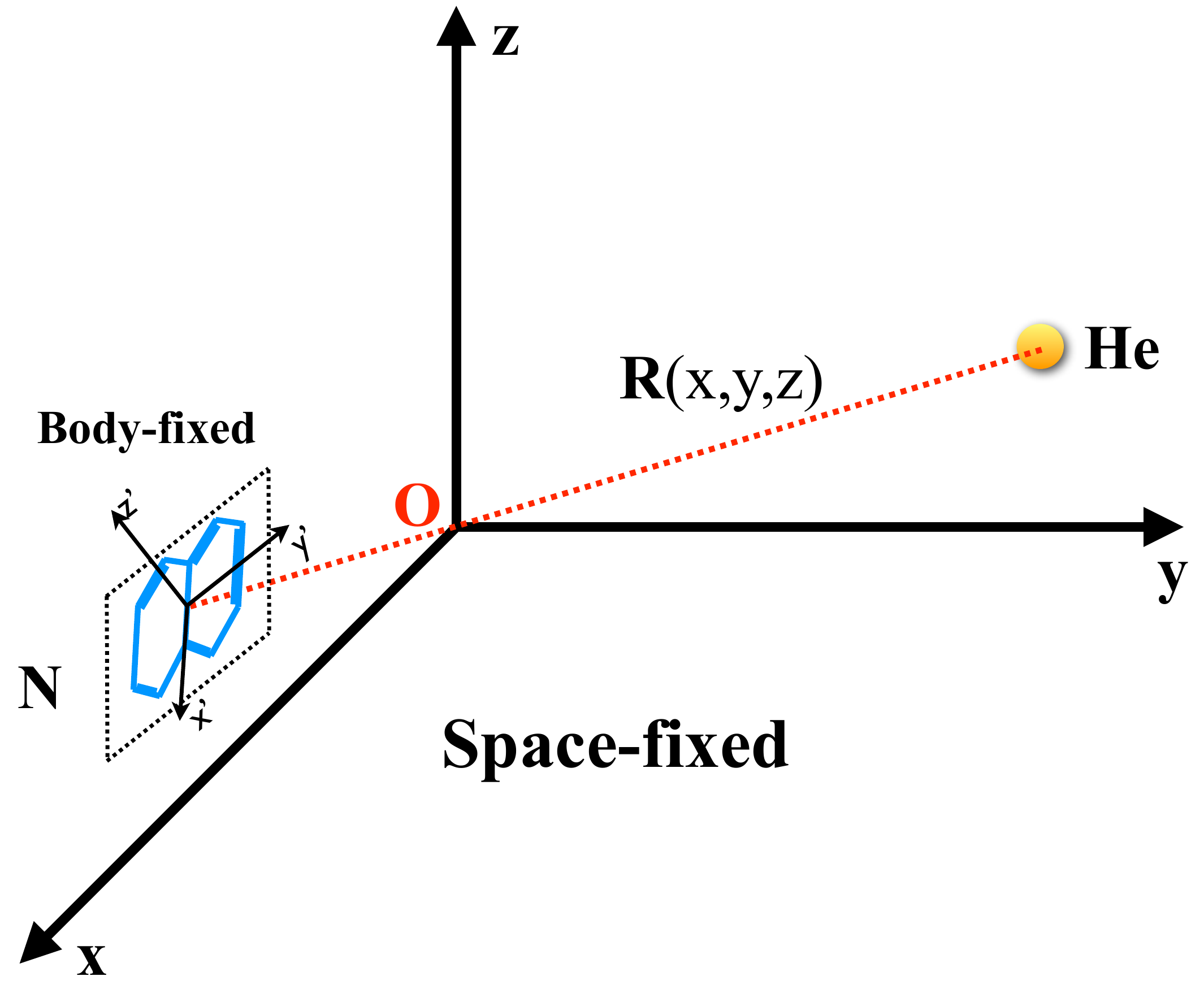}
\end{center}
\caption{The coordinate system for collisions between naphthalene (N) and helium at cold temperatures. The relative motion of N and He is described in a space-fixed frame ($x, y, z$) with the origin (O) located at the center of mass of the collision complex N-He. The inter-particle distance ${\bf R}$ is the position vector of He relative to the center of mass of the molecule N. The motion of N is described using a body-fixed frame ($x',y',z'$) with the origin positioned in the center of mass of N. The $x',y'$, and $z'$ axes correspond to the principal axes of the molecule.}
\label{coordinates}
\end{figure}

 Fig.~\ref{coordinates} demonstrates the coordinate system for collisions between naphthalene and helium at cold temperatures. The relative motion between naphthalene and helium is described in a space-fixed frame ($x, y, z$) with the origin (O) located at the center of mass of the collision complex N-He. The inter-particle distance ${\bf R}$ is the position vector of He relative to the center of mass of molecule N. The rotational motion of N is described using a body-fixed frame ($x',y',z'$) with the origin positioned at the center of mass of N. The $x',y'$, and $z'$ axes correspond to the principal axes of the molecule. The rotational constants of naphthalene molecules used in this calculation are: $A=0.10405$~cm$^{-1}$; $B=0.04113$~cm$^{-1}$; $C=0.02948$~cm$^{-1}$ \cite{Kabir}. The orientation of the molecule N with respect to the space-fixed system is described by Euler angles. When we integrate the equations of rotational motion of molecules, we transform Euler angles into quaternion parameters \cite{Evans1, Evans2, Rapaport} in order to eliminate the singularities in the equations of motion for the molecule. The approach is described in detail in the next section. The interaction potential surfaces between N and He are constructed based on the atom-bond pairwise additive representation developed by Pirani {\it et al} \cite{Pirani2,Pirani1}. 

\clearpage
\subsection{Collision dynamics}

The relative motion of naphthalene and helium is described in a space-fixed Cartesian coordinate system. The motion of particles is determined by numerically integrating a system of classical equations of motion:
\begin{equation}
\frac{d\bf{R}}{dt} = \frac{{\bf p}}{\mu} ;
\end{equation}
\begin{equation}
\frac{d{\bf p}}{dt} = {\bf F^{\rm SF}},
\end{equation}
where $t$ and ${\bf p}$ are the time and the momentum for the N$\--$He relative motion, respectively, and $\mu$ is the reduced mass. ${\bf F^{\rm SF}}$ is the interaction force between N and He, which is given by the first derivative of the interaction potentials $V({\bf R})$,
\begin{equation}
{\bf F^{\rm SF}} = -\nabla V({\bf R}).
\end{equation}

The orientation of naphthalene is described by quaternion parameters in a body-fixed frame, which are defined as follows:  
\begin{equation}
q_0 = \cos \(\frac{\theta}{2}\) \cos\frac{1}{2}\(\phi+\psi\),
\end{equation}
\begin{equation}
q_1 = \sin \(\frac{\theta}{2}\) \cos\frac{1}{2}\(\phi-\psi\),
\end{equation}
\begin{equation}
q_2 = \sin \(\frac{\theta}{2}\) \sin\frac{1}{2}\(\phi-\psi\),
\end{equation}
\begin{equation}
q_3 = \cos \(\frac{\theta}{2}\) \sin\frac{1}{2}\(\phi+\psi\),
\end{equation}
where $\theta$, $\phi$, and $\psi$ are Euler angles. The rotational motion of naphthalene molecules is determined by a set of second order differential equations \cite{Rapaport}:
\begin{align}
\left(\begin{array}{c}
\ddot q_0\\ \ddot q_1\\ \ddot q_2\\ \ddot q_3 \end{array}\right) = \frac{1}{2}
\left(\begin{array}{cccc}
-q_1&-q_2&-q_3&q_0\\
q_0&-q_3&q_2&q_1\\
q_3&q_0&-q_1&q_2\\
-q_2&q_1&q_0&q_3
\end{array}\right) 
\left(\begin{array}{c}
\dot \omega_x\\ \dot \omega_y\\ \dot \omega_z\\ -2\sum\dot q_m^2
\end{array}\right)
\label{q1}
\end{align}
where $m=0,1,2,3$; $\dot q_m$ and $\ddot q_m$ are the first and the second time derivatives of $q_m$, respectively; $\omega_{x,y,z}$ are the components of the angular velocity around the principal axes in the body-fixed frame, and $\dot \omega_{x,y,z}$ are their first derivatives with respect to $t$. In Equation~\ref{q1}, $\dot \omega_{x,y,z}$ can be obtained from the Euler's equations: 
\begin{equation}
\dot{\omega}_x = \frac{1}{I_x} \[N^{\rm BF}_x+(I_y-I_z)\cdot \omega_y\cdot \omega_z\] 
\end{equation}
\begin{equation}
\dot{\omega}_y = \frac{1}{I_y} \[N^{\rm BF}_y+(I_z-I_x)\cdot \omega_z\cdot \omega_x\] 
\end{equation}
\begin{equation}
\dot{\omega}_z = \frac{1}{I_z} \[N^{\rm BF}_z+(I_x-I_y)\cdot \omega_x\cdot \omega_y\],
\end{equation}
where $I_{x,y,z}$ are the principal moments of inertia of the molecule and $N^{\rm BF}_{x,y,z}$ are the components of the torque in the body-fixed frame.  We can calculate $N^{\rm BF}_{x,y,z}$ by transforming the torque in the space-fixed frame $N^{\rm SF}_{x,y,z}$ with a rotational matrix $\mathcal{A}$ \cite{Evans1, Evans2, Rapaport}
\begin{equation}
{\bf N^{\rm BF}} = \mathcal{A} {\bf N^{\rm SF}},
\end{equation}
defined as follows
\begin{align}
\mathcal{A} = 
\left(\begin{array}{ccc}
\frac{1}{2}-q^2_2-q^2_3 & q_1q_2+q_0q_3 & q_1q_3-q_0q_2\\
q_1q_2-q_0q_3&\frac{1}{2}-q_1^2-q_3^2&q_2q_3+q_0q_1\\
q_1q_3+q_0q_2&q_2q_3-q_0q_1&\frac{1}{2}-q_1^2-q_2^2
\end{array}\right).
\end{align}
The torque is exerted on the center-of-mass of the molecule due to the interaction between the molecule and the helium atom. ${\bf N^{\rm SF}}$ is then given by:
\begin{equation}
{\bf N^{\rm SF}} = {\bf R} \times (-{\bf F}^{\rm {SF}}).
\end{equation}

In this calculation, we use the fourth-order Runge-Kutta method to solve coupled differential equations for both translational and rotational motions of the particles. The time stepsize for the integration is $5\times10^{-3}$~ps. For low collision energies, e.g., $\leq 1$~$\rm{cm}^{-1}$, the stepsize is chosen to be $10^{-5}$~ps. The total energy during collisions can be conserved up to three digits and the structure of naphthalene remains rigid throughout the simulation. The trajectories are terminated at large inter-particle separations ($R_{\rm final}$) where the interaction potentials can be ignored.  

In this paper, we calculate two physical quantities: (i) the total cross sections for collisions between naphthalene and helium; (ii) the lifetime of the quasi-bound collision complex N$-$He$^*$. We choose $R_{\rm final}$ to be $50$~\AA~in the calculation of (i) and $20$~\AA~for (ii). The lifetime of the collision complex N$-$He$^*$ is determined as the time between the first and the last turning points. The rotational energy of naphthalene can be calculated from the final principal rotational velocities of naphthalene $\omega'_{x,y,z}$, which are given by the following equation
\begin{align}
\left(\begin{array}{c}
\omega_x'\\ \omega_y'\\ \omega_z'\\ 0 \end{array}\right) = 2
\left(\begin{array}{cccc}
-q_1&q_0&q_3&-q_2\\
-q_2&-q_3&q_0&q_1\\
-q_3&q_2&-q_1&q_0\\
q_0&q_1&q_2&q_3
\end{array}\right) 
\left(\begin{array}{c}
\dot q_0\\ \dot q_1\\ \dot q_2\\ \dot q_3
\end{array}\right).
\end{align}

The trajectories for numerical calculations are generated with the initial conditions chosen as follows. The initial rotational energy of the molecule $E_{\rm rot}$ is determined by diagonalizing the matrix constructed with the hamiltonian for the rotational motion of a free asymmetric-top rigid rotor. We randomly sample the Euler angles and the magnitudes of $\omega_{x,y,z}$ with the constraint $I_x\omega_x^2+I_y\omega_y^2+I_z\omega_z^2 = E_{\rm rot}$ to ensure that the trajectories cover the entire phase space \cite{Pattengill}.  The impact parameter $b_l$ is chosen quantum mechanically \cite{Pattengill}:
\begin{equation}
b_l = \sqrt{l(l+1)} \hbar/P^{\rm SF},
\end{equation}
where $l$ is the quantum number for the angular momentum describing the relative rotation between naphthalene and helium. For the calculation of (i), the value of the largest impact parameter $b_l^{\rm max}$ is determined so that no trajectories have rotational energy transfer larger than the smallest energy gap between the rotational energy levels of naphthalene. For (ii), $b_l^{\rm max}$ is chosen to be large enough so that the lifetime of  N$-$He$^*$ obtained from the calculations of the trajectories with $b_l > b_l^{\rm max}$ is zero. We carry out calculations of $4\times10^5$ trajectories for each collisional energy and the results were converged up to two digits. The total number of trajectories in this calculation is on the order of $10^7$. 

\section{Results}

\begin{figure}[ht]
\begin{center}
\includegraphics[scale=0.5]{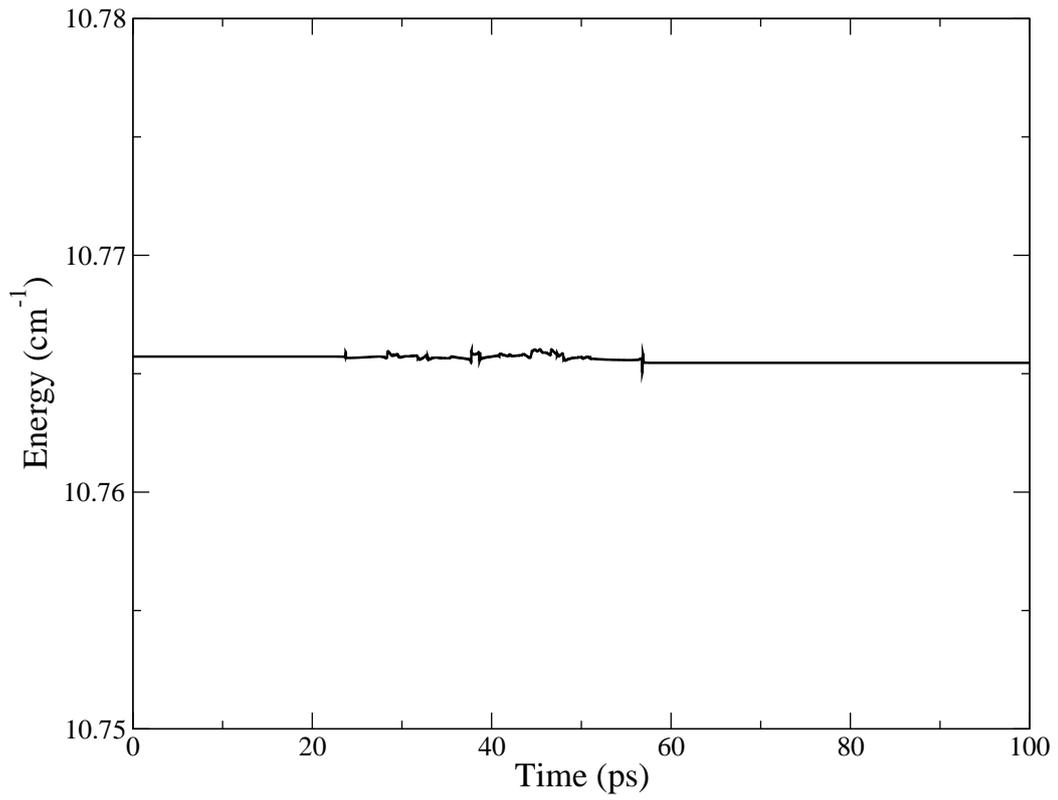}
\end{center}
\caption{ The total energy of the N-He collision complex as a function of time at the collision energy of 6~cm$^{-1}$. The energy of the system is intrinsically conserved up to within three digits.}
\label{EC}
\end{figure}

\begin{figure}[ht]
\begin{center}
\includegraphics[scale=0.5]{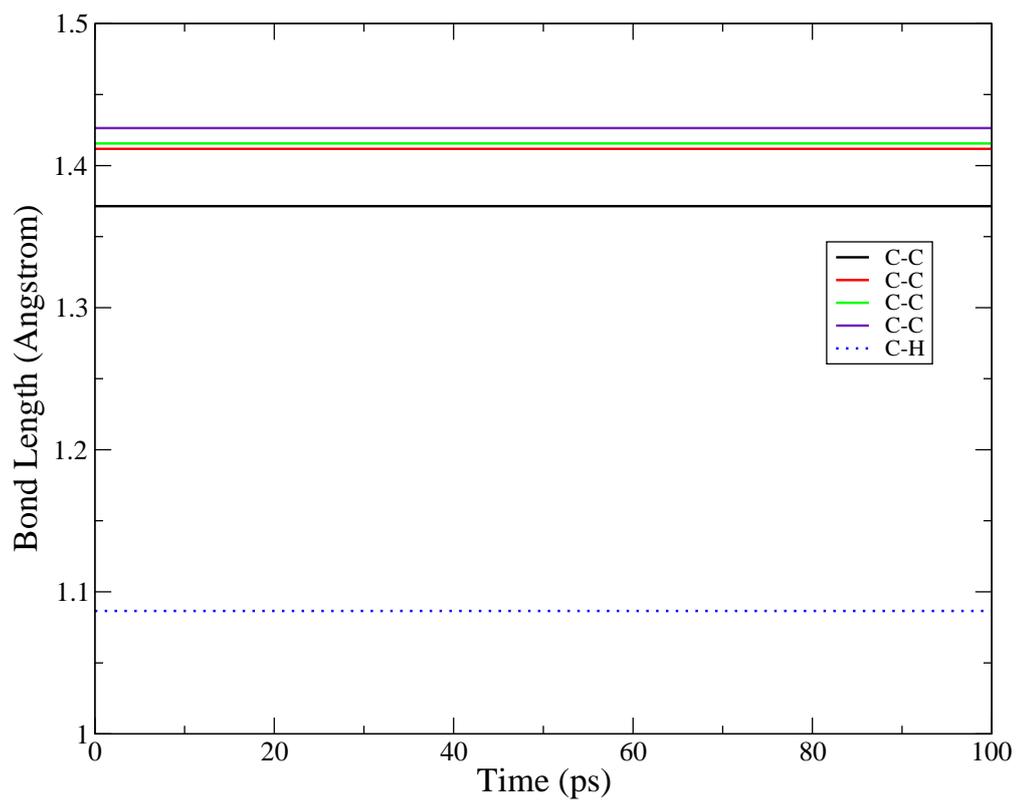}
\end{center}
\caption{ The C$\--$C and C$\--$H bond lengths of naphthalene as functions of time during the collision process. The rigidity of Naphthalene is fully preserved.}
\label{BL}
\end{figure}

In classical trajectory studies of collision dynamics of molecules, due to the accumulation of numerical errors, the conservation of the total energy of the collision complex and the rigidity of the molecules are generally difficult to maintain. In order to check the validity of our calculations, we calculate the total energy of the N-He collision complex and the C$\--$C and C$\--$H bond lengths of naphthalene as functions of time, as shown in Fig.~\ref{EC} and Fig.~\ref{BL}, respectively. Despite the fluctuations of the total energy in the regime of short-range interactions between naphthalene and helium, the total energy of the collision complex is intrinsically conserved to within three digits. The structure of molecules remains rigid throughout the simulation.

\begin{figure}[ht]
\begin{center}
\includegraphics[scale=0.5]{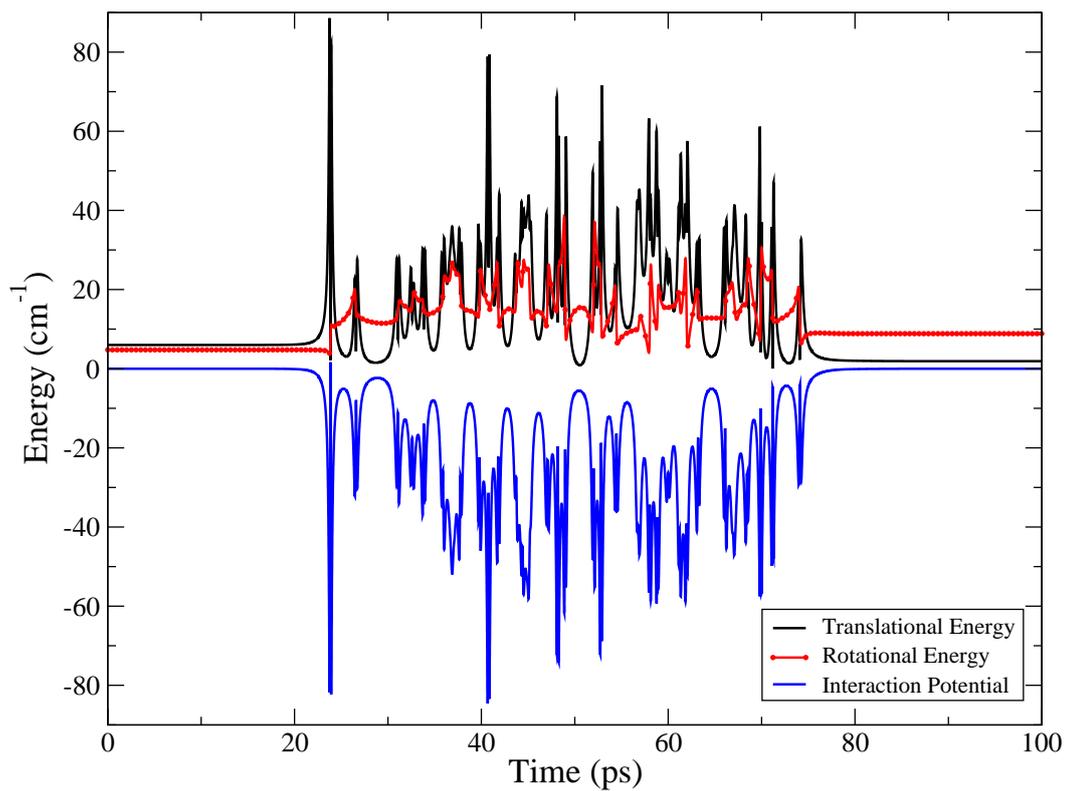}
\end{center}
\caption{The translational energy (black curve), rotational energy (red dotted curve), and interaction potential energy (blue curve) as functions of time during a collision process between naphthalene and helium.  }
\label{ET}
\end{figure}

\begin{figure}[ht]
\begin{center}
\includegraphics[scale=0.5]{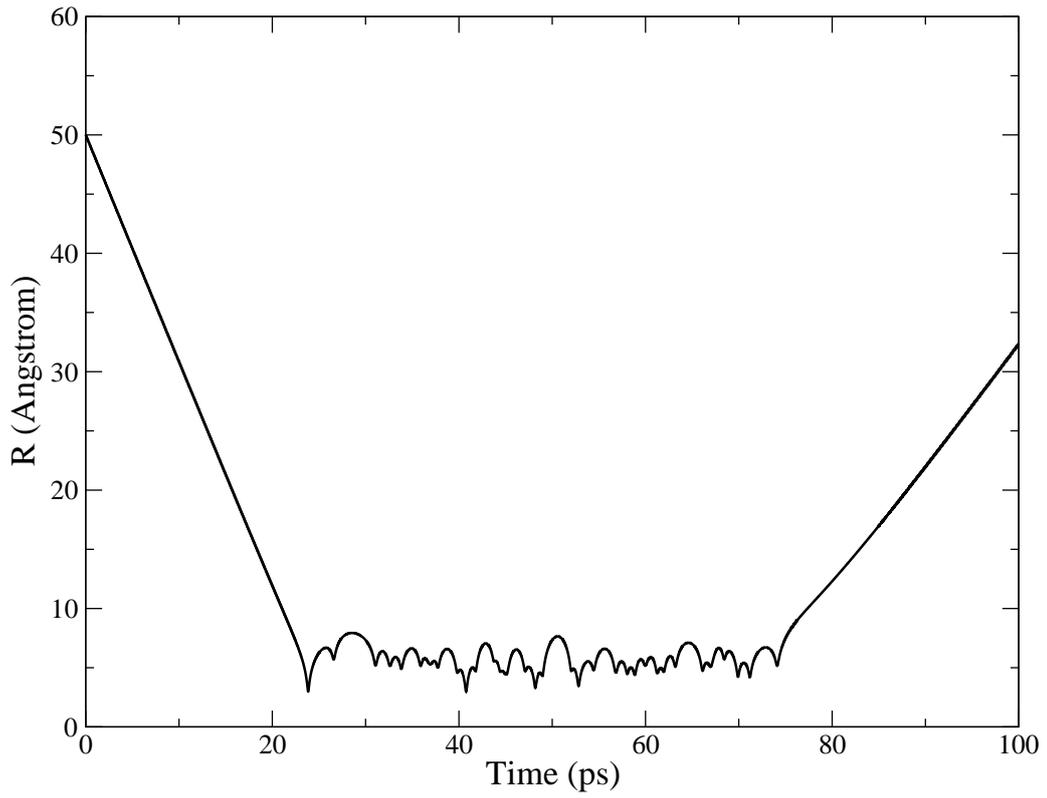}
\end{center}
\caption{The inter-particle distance between naphthalene and helium as a function of time. The helium atom is trapped near the collision center from 24.00~ps to 74.25~ps.}
\label{2be-traj}
\end{figure}

\begin{figure}[ht]
\begin{center}
\includegraphics[scale=0.45]{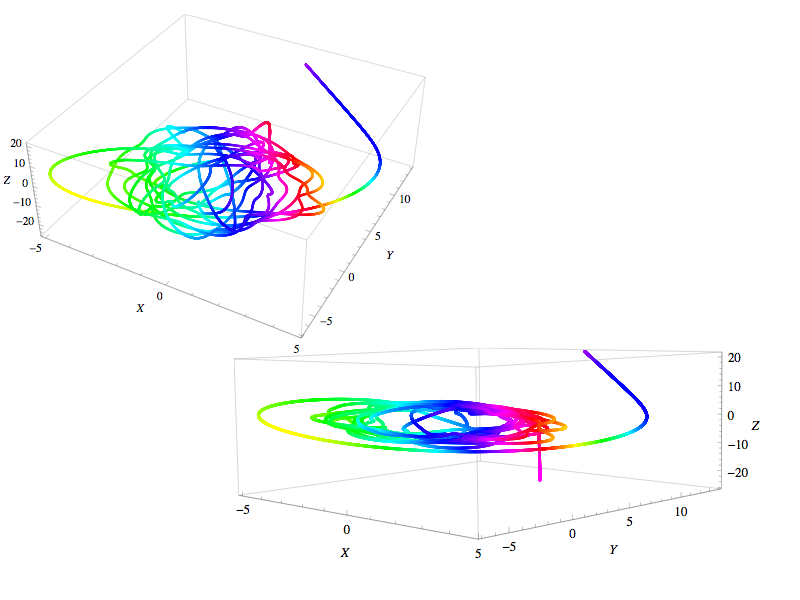}
\end{center}
\caption{The $z$ coordinates of the relative motion between helium and naphthalene as a function of its $x$ and $y$ coordinates for a long classical trajectory.}
\label{traj}
\end{figure}

At low temperatures, collisions between inert gas atoms and a complex molecule may induce energy transfer between translational and internal degrees of freedom. As a result, the atom may be trapped and move around the molecule for a long time leading to the formation of a long-lived quasi-bound collision complex. We observe some trajectories in our calculations where the exchange between relative translational energy and rotational energy of the molecule is very active, as demonstrated in Fig.~\ref{ET}. In this trajectory, the translational energy (black curve) oscillates significantly due to the interaction between naphthalene and helium and there are many rotational excitations and de-excitations of naphthalene (red dotted curve) during the collision process. At the end of the trajectory, the rotational energy of naphthalene is elevated from 4.75~cm$^{-1}$ to 8.85~cm$^{-1}$; correspondingly, the translational energy decreases from 6.00~cm$^{-1}$ to 1.89~cm$^{-1}$. Fig.~\ref{2be-traj} shows for the same trajectory the inter-particle distance between naphthalene and helium as a function of time. The helium atom is trapped near the collision center from 24.00~ps to 74.25~ps where a long-lived collisional complex is formed. The lifetime of the complex in this trajectory is about 50.49~ps, which is determined as the time the particles spend from the first to the last turning points, as shown in Fig.~\ref{ET}. Fig.~\ref{traj} displays for the same trajectory the $z$ coordinates of the relative motion between helium and naphthaline as a function of its $x$ and $y$ coordinates. This graph depicts the trajectory in three dimensions. It demonstrates an orbital feature of the motion of the particles.

\begin{figure}[ht]
\begin{center}
\includegraphics[scale=0.5]{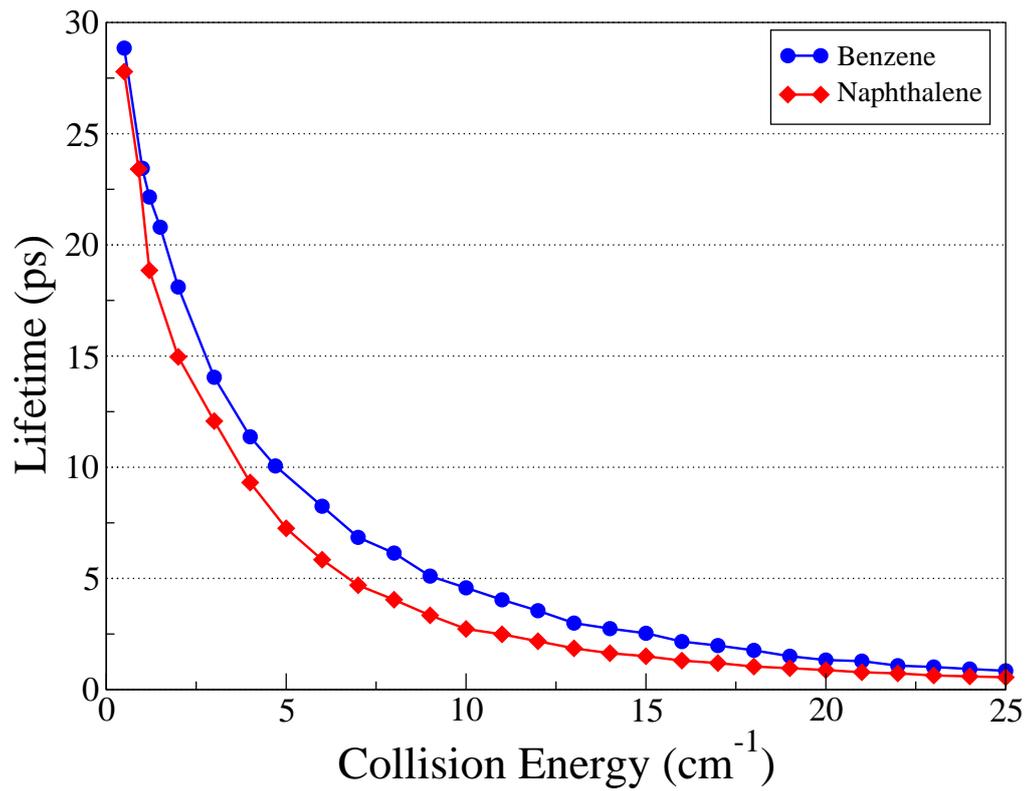}
\end{center}
\caption{The average lifetime of the quasi-bound collision complex of N$-$He$^*$ (red curve) and B$-$He$^*$ (blue curve) as functions of collision energy. }
\label{LT}
\end{figure}

\begin{figure}[ht]
\begin{center}
\includegraphics[scale=0.7]{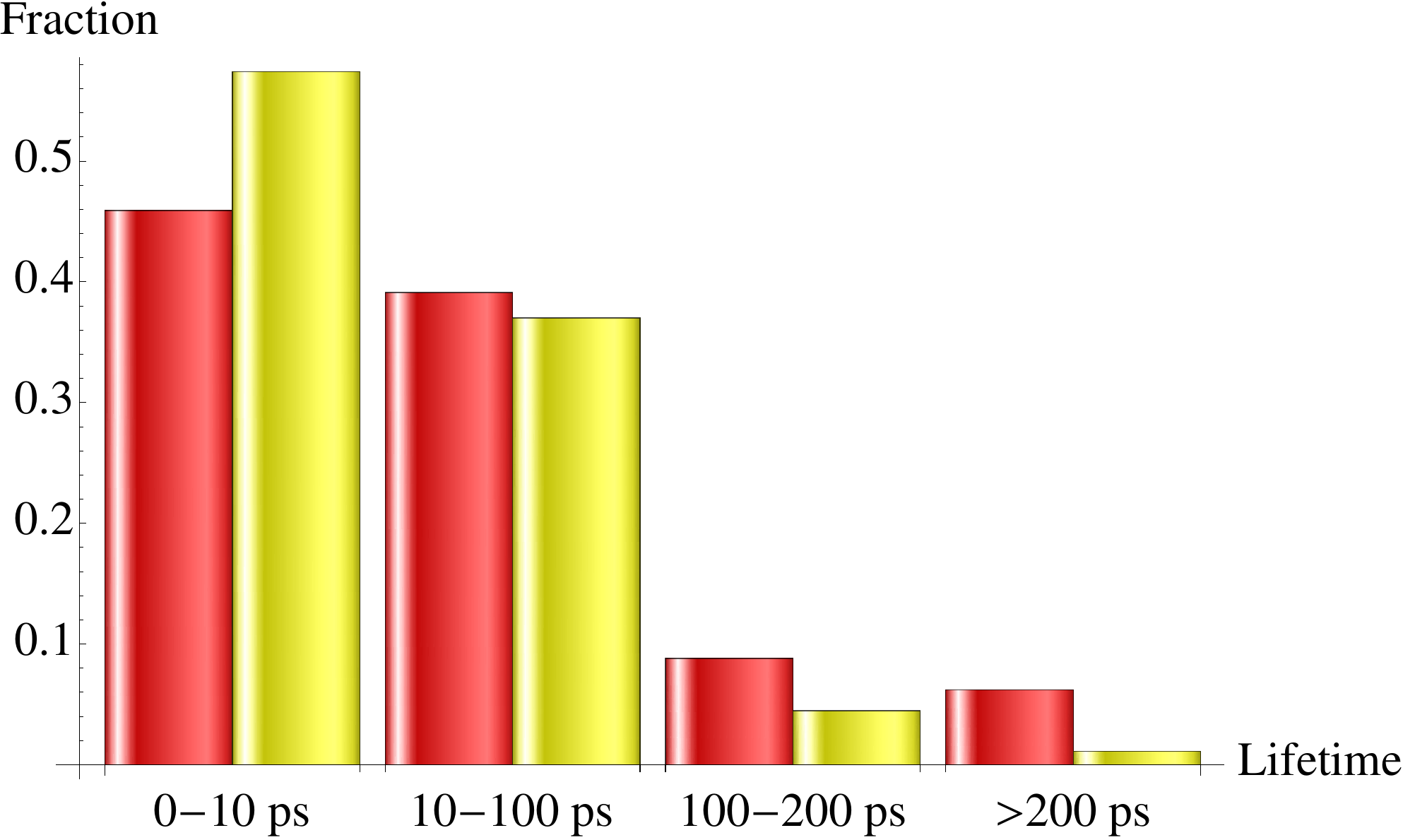}
\end{center}
\caption{The fraction of classical trajectories at a particular lifetime interval as a function of the lifetime of the B$-$He$^*$ (red) and N$-$He$^*$ (yellow) complexes at a collision energy of 2~cm$^{-1}$  for $J = 0$.  Naphthalene has a bigger fraction of short trajectories (lifetime < 10~ps) than benzene, while benzene has a larger fraction of long trajectories. This difference becomes more significant as the lifetime of the collision complex increases.}
\label{com}
\end{figure}

In order to explore the role of the molecular shape in determining cold collisions, we calculate the average lifetime of the quasi-bound collision complexes B$-$He$^*$ and N$-$He$^*$  as a function of the collision energies, as shown in Fig.~\ref{LT} by red curves and blue curves, respectively. Each point in this graph is the lifetime averaged over all trajectories with rotational energies of molecules below 20~cm$^{-1}$ according to the Boltzmann distribution. Both curves display a decay of the lifetime as the collision energy increases. However, contrary to expectation, the lifetime of the B-He complex is longer than that of N-He even though benzene is lighter than naphthalene. Naphthalene is an asymmetric top while benzene is a symmetric top. We have established that the symmetry of benzene plays a twofold role in increasing the lifetime of the collision complex. First, it enhances the probability of the occurrence of long trajectories. Based on our observations, long trajectories mainly occur when helium approaches the molecules near their edges. For a certain impact parameter which may lead to a long-lived collision complex, benzene allows for more approach directions than naphthalene that generate long trajectories. As a result, benzene molecules have a larger fraction of long trajectories than naphthalene. This mechanism is clearly demonstrated in Fig.~\ref{com}, which shows the fraction of classical trajectories at a particular lifetime interval as a function of the lifetime of the B$-$He$^*$ (red) and N$-$He$^*$ (yellow) complexes at a collision energy of 2~cm$^{-1}$  for $J = 0$. Naphthalene has a bigger fraction of short trajectories (lifetime < 10~ps) than benzene, while benzene has a larger fraction of long trajectories. This difference becomes more significant as the lifetime of the collision complex increases. Second, molecular symmetry may enhance the number of rotational periods in a molecule-helium complex, which leads to trapping of helium for a very long time. To illustrate this, we compare the lifetime of the longest trajectory for benzene and naphthalene. The longest lifetime for B$-$He$^*$ in our calculation is 1543~ps, while for N$-$He$^*$, it is only 592~ps. These results suggest that the geometry of molecules plays a significant role in determining the dynamics of cold collisions. 

The long-lived quasi-bound collision complex, as demonstrated in Fig.~\ref{ET}, \ref{2be-traj}, and \ref{traj}, could result in a stable molecule-He complex if the lifetime of the complex is long enough to allow subsequent collisions with helium. Here, we average the lifetime of the  N$-$He$^*$ and B$-$He$^*$ complexes over the collision energy according to the Maxwell-Boltzmann distribution and compare it with the mean time a helium atom needs to collide with a naphthalene molecule $\tau_{\rm {He}}$. The mean lifetime of the N$-$He$^*$ and B$-$He$^*$ complexes is 7.9~ps and 9.9~ps, respectively. $\tau_{\rm {He}}$ can be estimated as an inverse of the number of collisions $\mathcal{N}$ one naphthalene molecule experiences in a helium gaseous bath per second, which is given by
\begin{equation}
\mathcal{N} = N_{\rm He} \sigma_{\rm N-He} \sqrt{\frac{8k_{\rm B}T}{\pi \mu_{\rm N-He}}}.
\label{time}
\end{equation}
In Eq.~\ref{time}, $N_{\rm He} = 4 \times 10^{17}$~cm$^{-3}$ is the density of helium atoms used in the experiments \cite{Patterson2}, $\sigma_{\rm N-He} \sim 10^{-13}$~cm$^{2}$ is the total thermal cross section for collisions between naphthalene and helium \cite{Patterson2}, $k_{\rm B}$ is the Boltzmann factor, and $T = 6.5$~K is the temperature of the naphthalene-helium gaseous mixture. Using Eq.~\ref{time}, we obtain $\mathcal{N} \sim 7.24 \times 10^8 \rm~s^{-1}$ and correspondingly, $\tau_{\rm {He}} \sim1380~\rm ps$. This time is two orders of magnitude longer than the calculated average lifetime of the complexes. The time a helium atom needs to collide with the quasi-bound collision complex should be even longer than $\tau_{\rm {He}}$.  This indicates that the mean lifetime of the quasi-bound collision complex N$\--$He* is too short to allow subsequent collisions with a second He to form a stable molecular complex. As a consequence, no clustering occurs for naphthalene in a low-temperature He gas. This suggests that the density of the helium gaseous bath used in the buffer-gas cooling experiment could be increased up to  $10^{19}$~cm$^{-3}$, which may decrease the loss of the produced molecules during the cooling process and enhance the efficiency of cooling.  

We also calculate the thermal total cross section for collisions between naphthalene and helium. The cross section is  $1.19\times10^{-13}$~cm$^2$, which is consistent with experimental observation $\sigma_{\rm N-He} \sim 10^{-13}$~cm$^{2}$  \cite{Patterson2}.

\section{Summary}

We have introduced a method for classical trajectory calculations of collisions between atoms and large rigid asymmetric-top molecules. We explore the collision dynamics between helium and large polyatomic molecules at low temperatures. In particular, we investigate the formation of naphthalene-helium and benzene-helium complexes in buffer-gas cooling experiments. Our finding provides an explanation for the experimental observation. Clustering would occur if helium atoms formed long-lived quasi-bound collision complexes with naphthalene molecules.  Subsequent collisions with helium would result in the formation of stable naphthalene-helium complexes. Our calculations show that the mean lifetime of the collision complex N$\--$He* is too short to allow subsequent collisions. As a consequence, no clustering occurs under the experimental condition. In the experiment, the density of helium atoms is about $4 \times 10^{17}$~cm$^{-3}$. Our calculations show that this density could be increased to $10^{19}$~cm$^{-3}$ for the buffer-gas cooling of naphthalene molecules. In buffer-gas cooling, the density of helium determines the time scale for the thermalization of translational degrees of freedom. This time scale should be short enough to guarantee that the target molecules can be cooled to the desirable temperature before they are lost on the wall of the cell. Increasing the helium density therefore could potentially improve the efficiency of the production of cold molecules. In addition, we calculate the lifetime of the quasi-bound complex for collisions between benzene and helium at the temperature of 6.5~K. Contrary to the expectation, the mean lifetime of B$\--$He* is longer than that of N$\--$He*. This is due to the symmetric structure of benzene. Our results indicate that the shape of molecules is important in determining the dynamics of cold collisions, in some cases, even more crucial than the number of accessible degrees of freedom. This suggests that by selecting molecules with suitable shape for buffer-gas cooling, one could cool molecules with a very large number of degrees of freedom. 

\section{Acknowledgement}
The work was supported by NSERC of Canada.

\newpage

\newpage

\newpage

\newpage

\newpage

\newpage

\newpage

\newpage

\newpage

\end{document}